\documentclass[sigconf, nonacm]{acmart}

\definecolor{shmgreen}{RGB}{126, 166, 61}

\AtBeginDocument{%
  }

\setcopyright{none}
\acmConference[Under Review]{}{}

\usepackage{booktabs}
\usepackage{array}
\usepackage{graphicx}
\usepackage{tabularx}
\usepackage{threeparttable}

\begin{document}

\title{Mylar: Exploring Self-Curation as a Reflexive Creative Practice}


\author{Kate Chier}
\affiliation{%
  \institution{University of California, Berkeley}
  \city{Berkeley}
  \state{CA}
  \country{USA}}
\email{katechier@berkeley.edu}

\author{Shm Garanganao Almeda}
\affiliation{%
  \institution{University of California, Berkeley}
  \city{Berkeley}
  \state{CA}
  \country{USA}}
\email{shm.almeda@berkeley.edu}

\author{Bjoern Hartmann}
\affiliation{%
 \institution{University of California, Berkeley}
  \city{Berkeley}
  \state{CA}
  \country{USA}}
\email{bjoern@berkeley.edu}

\renewcommand{\shortauthors}{Chier et al.}


\begin{abstract}
The container holding an artist’s work can itself be just as expressive as the work it presents. We investigate self-curation as a creative practice, asking how artists explore and express their identity through the design of their own artwork presentations. Drawing on a study of artists’ existing presentation practices, we developed Mylar, a malleable art presentation system that lets artists rapidly generate and compare provisional portfolios, seeing their work anew through different montages. We observed how 12 artists articulated, probed, negotiated, and foraged in Mylar, using rapidly generated portfolios as epistemic artifacts for reflection. We characterize self-curation as a reflexive and plural practice, in which artists use presentation design to shape the legibility of their self and their work and reflexively correspond with provisional presentations to reshape their curatorial intentions. We share design implications for expressivity in online art platforms and for malleable prototyping as a formative design method.
\end{abstract}

\ccsdesc[500]{Human-centered computing~Empirical studies in HCI}
\ccsdesc[300]{Human-centered computing~Interactive systems and tools}
\ccsdesc[300]{Human-centered computing~HCI theory, concepts and models}

\keywords{self-curation, artist portfolios, creativity support tools, malleable interfaces, reflective practice}

\begin{teaserfigure}
 \includegraphics[width=\textwidth]{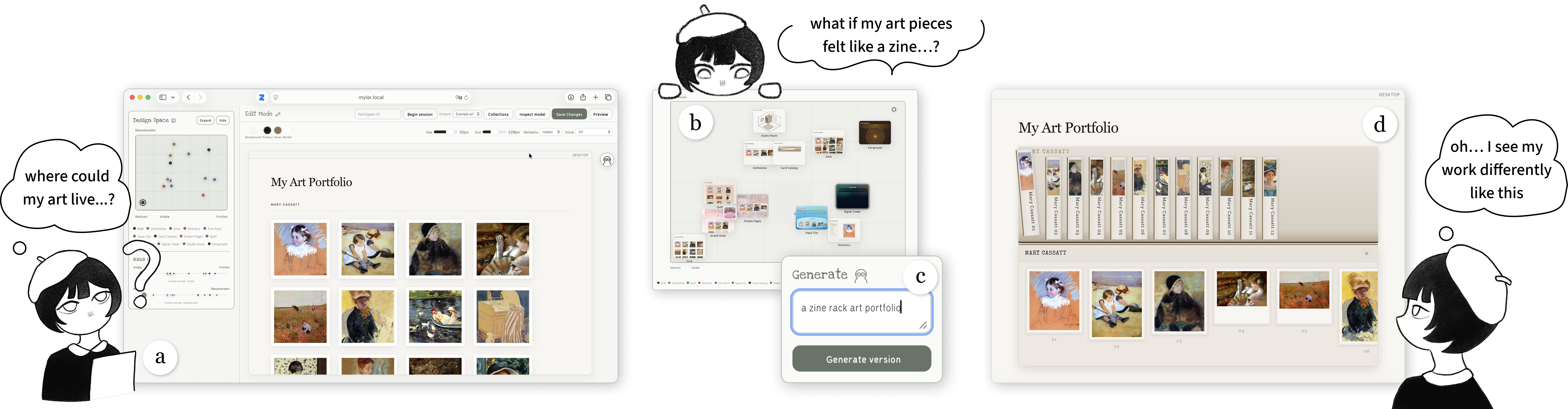}
  \caption{Mylar explores self-curation as a creative and reflective practice. An artist moves through (\textbf{a}) the Mylar workspace, (\textbf{b}) the expanded design space, and (\textbf{c}) the generation panel to produce a (\textbf{d}) generated portfolio, using presentation to reconsider relationships within the work and what it might mean.}
  \Description{Four-panel illustration of an artist exploring and generating portfolio presentations in Mylar. From left to right, (a) the workspace shows a grid-based portfolio beside a compact two-dimensional design space containing several portfolio directions; (b) the expanded design space shows thumbnail previews distributed across the map; (c) a generation panel contains the prompt ``a zine rack art portfolio''; and (d) the resulting portfolio presents the same artworks in a zine-rack-inspired interface, with works displayed as upright publications above a conventional image collection. Illustrated figures frame the sequence, moving from wondering where their art could live to noticing that the new presentation makes them see their work differently.}
  \label{fig:teaser}
\end{teaserfigure}


\maketitle

\section{Introduction}
\begin{quote}
``The place in which the work is shown imprints and marks this work.''
\begin{flushright}
    ---Daniel Buren, \textit{The Function of the Museum} \cite{museum_function_2019}
\end{flushright}
\end{quote}

An artwork carries an artist's time, skill, and creative expression---and that same artistry persists in the containers artists choose or create to present their work. Curatorial scholarship has long understood presentation as an interpretive practice, where the arrangement and juxtaposition of artworks can produce new meanings \cite{curation_2013, becker_art_2008}. 
In self-curation---the practice by which artists select and present a collection of their artwork---presentation becomes a \textit{self-reflexive }interpretive practice, through which artists can extend their expression. In composing arrangements of their artwork, artists become curators of their own practice, extending their artistic vision beyond each individual piece and into shaping an audience's encounters with their collection. 

Creativity support tool (CST) research in HCI has long attended to the processes by which creative artifacts are produced \cite{landscape_2019, cst_2007}. A growing body of critical perspectives signals a return to more collective and pluralistic lenses on creativity support~\cite{nakakoji_framework_1999,cse_2025}, calling for researchers and system designers to expand their focus beyond the production of a particular artifact ~\cite{beyond_productivity_2025, beyond_artifact_2023}. HCI researchers have also investigated and developed tools to support expert curation of historical \& cultural art (e.g., for museums and galleries), and tools for the digital presentation of historical collections~\cite{artographer_2026, collections_2023}. But today, countless living artists are sharing and presenting their work to global online communities~\cite{crain_share_2017}.  Artists seeking out the support of a creative community~\cite{mosaic_2017, campbell_thousands_2016-1} build sites and social media collections to develop their online presence and practice---in spite of the labor and risks of online creative distribution, and the lack of tools and structures designed to support them through it~\cite{cse_2025, von_davier_looking_2025, fiesler_understanding_2015}.

Interested in the design of systems to support self-curation, we first set out to understand self-curation using existing forms of design ideation. We conducted a formative study, sharing example sites and sketching speculative presentation concepts with 7 artists. We learned how artists' presentation needs continually shift alongside their practice, and how intermediate design representations---example sites and sketches---helped artists surface their design needs and articulate a variety of divergent concepts and considerations for presenting their body of work. Recent advances in code generation afford the rapid creation of digital prototypes\cite{pail_2025}. This shaped our vision of a system that would help artists  consider their work across a variety of presentations---to flexibly envision, prototype, and evaluate a space of possible presentation designs. 
 
We ask: \textbf{How might rapidly reimagining the presentation of their work shape artists' practices of self-curation?}

To investigate this question, we drew from malleable design principles~\cite{generative_interfaces_2025, malleable_overview_2025} to develop Mylar, a system that supports presentation design ideation through provisional artifacts. Artists can sketch portfolio possibilities through language, visual references, and structured design dimensions, then directly reshape the resulting portfolios, and reuse elements across directions. We sought to explore the feasibility and functionality of rapidly generated prototypes as \textit{epistemic artifacts}~\cite{epistemic_2024}: materials through which artists test and revise their own understanding of what they want their work to become.

We conducted a discovery-oriented evaluation, a kind of design probe study, with 12 artists \cite{systems_2026}, who each used Mylar to generate and manipulate provisional presentations of their own artwork collections. Across these sessions, we observed how artists used these provisional portfolios to articulate, probe, negotiate, and forage---rendering different facets of their work, practice, and artistic identity legible in turn. We consider the role of these generated prototypes beyond the designs artists ultimately keep---how they can surface possibilities and serve as materials through which artists recognize, question, and reshape their curatorial intentions.

Through analysis of artists' engagements with Mylar---and the ways they created and used materials in Mylar to engage with us as researchers---we co-construct a deeper understanding of the ways artists can extend their expressive practice into designing the way their work is encountered, developing an account of self-curation as a medium. We characterize the reflexivity of this medium: how artists shape the legibility of their self and their work through presentation design, and how their encounters with provisional presentations can clarify, unsettle, and reshape self-curatorial intentions. From this account, we discuss implications and provocations for the design of online art platforms: rather than prescribing a single, standardized, static portfolio format, how might we instead support each artist's expressivity and autonomy over their online presences, empowering them to flexibly present pluralistic, evolving accounts of their selves and their practice? We also draw on our experiences building and using Mylar for design ideation with this community, to discuss opportunities---and potential pitfalls---of using malleable prototyping as a method for formative design ideation.

\section{Related Work}

\subsection{Supporting Art Curation and Presentation in HCI}
There is a long lineage of systems designed to support curation and engagement with art collections, especially systems for expert museum and gallery curators~\cite{tuscher_nodes_2025, ohm_collection_2023, saleh_large-scale_2016}. 
Art researcher support tools help art historians explore, analyze, transform, and curate cultural collections~\cite{kaplan2016visual, leonard_pleonard212pix-plot_2026}. Interactive and flexibly configurable visualizations can help art researchers see a collection in different arrangements and through different lenses, placing the works in new conversations with one another, and thereby generating new ways of understanding them. Interactive digital presentations can also make collections more accessible or engaging, promoting public engagement with historical artwork~\cite{artographer_2026, kobeisse_moving_2024, kortbek_communicating_2008}. von Davier argues for the design of deeper and more meaningful interactions with digital art presentations~\cite{von_davier_designing_2023} and that such work is key in countering the ongoing ``contentification'' of online media~\cite{von_davier_looking_2025}.

While using computational interventions to support curation and interpretation of historical art collections is well-established, there is a relative dearth of systems for \textit{self}-curation---designed to support living artists in the relation, interpretation, and presentation of their own artworks.

\subsection{Art Presentation as  Creative Practice}
Artists' online presentations are more than neutral containers for holding artwork; they frame how it is encountered, interpreted, and valued across various audiences and platforms \cite{intermediaries_2025}. This expressive role has long been visible in authored personal homepages, through which individuals exercised substantial control over their online self-presentation~\cite{home_pages_1999, presentation_2002}. Recently, as online presence has become increasingly mediated through social media platforms and templated website builders, self-presentation is also shaped by platform-defined formats, visibility logics, and predefined design structures~\cite{platformization_2015, template_2024, creep_2025}. The act of presentation therefore unfolds within a broader online art ecosystem, where visibility, circulation, and platform participation extend beyond the production of individual artifacts \cite{cse_2025}. Prior systems have begun to support artists in constructing and distributing digital platforms, including AI-assisted tools for translating creative visions into interactive portfolio designs \cite{portfoliomentor_2023} and platforms that combine portfolio presentation, audience engagement, and commerce \cite{commerce_2023}. However, these approaches largely treat the portfolio as something to construct and communicate through, rather than a provisional surface that can be remade, revisited, and used to discover how an artist wants their work to live online.

\subsection{Prototyping as Epistemic Practice}
Creative artifacts can function not only as outcomes of design, but as materials through which intentions are developed and revised. Schön describes design as a ``reflective conversation'' with materials, while work on epistemic action similarly shows how external representations can make problems easier to think through~\cite{reflective_1983, distinguishing_1994}. Prototypes extend this role by allowing designers to explore and express questions through provisional artifacts rather than merely approximate finished systems~\cite{prototype_1997}. By making possibilities experimentally available, prototypes can become generative surfaces for reflection: creators encounter consequences they could not fully anticipate, respond to the ``backtalk'' of the material, and gradually develop their goals through the process of making~\cite{experience_2000, anatomy_2008, epistemic_2024, quickpose_2023}.

Multiple alternatives can extend this reflective role of prototyping by giving creators a broader space of possibilities to compare and respond to. Prior research in parallel and comparative prototyping has shown that alternatives can support comparative reasoning and situated exploration \cite{juxtapose_2008}, sharpen critique and evaluation \cite{right_design_2006}, and foster greater divergence while reducing friction \cite{parallel_prototyping_2011}. Generative AI widens this space of possibility by allowing alternatives to be produced in abundance. Recent systems have treated generated outputs not as isolated endpoints, but as material for structured comparison and sensemaking across a design space \cite{dreamsheets_2024}. At the same time, work on LLM-supported design warns that generation workflows can collapse this broader space into a single point solution, obscuring alternatives that may better reflect the user's goals \cite{pail_2025}.

\subsection{Navigating Generative Design Spaces}
Generative systems can collapse the distance between intention and artifact, even when intention is still taking shape. Creative intentions may begin vague and incomplete, leaving users uncertain about what they want, how to express it, or what the system makes possible \cite{gulf_2024, kreminski_endless_2026}. Prior work on mixed-initiative co-creativity similarly treats intent as something that can be surfaced and refined through interaction, rather than captured in a single prompt \cite{intent_2024}. This outlook shifts the role of AI support away from simply helping designers efficiently realize their intent, towards \textit{lensing it}---helping designers understand, shape, and amplify their creative intents, ideally without displacing or homogenizing them. 

When intentions are still forming, a single answer may be less useful than a landscape of possibilities. Recent systems have begun to structure generative outputs through spreadsheet-based prompt spaces for comparing variations \cite{dreamsheets_2024}, explicitly modeled design spaces that organize dimensions \cite{luminate_2024}, and multilevel representations that let users move between broader and more detailed views of generated outputs \cite{sensescape_2023}. Beyond using generation as a means for production, these systems contextualize generation to help make design space visible and navigable.

Exploration does not always resolve into a finished artifact, and can leave an interface deliberately open to further change. Prior work has sought to expose computational power through more direct forms of interaction, allowing artists to shape procedural behavior without having to author it in code \cite{para_2025}. More recent systems extend this idea to the web, from natural-language restyling of existing pages \cite{stylette_2022}, to end-user control over interface content, composition, and layout \cite{malleable_overview_2025}, to generated interfaces that continue changing as users' needs evolve \cite{generative_interfaces_2025}.

\section{Formative Study}
To ground our system design goals, we first set out to understand self-curation using existing forms of design ideation. We conducted a formative study, interviewing and conducting design ideation activities with 7 artists who actively share creative work online. Initial findings from this study were reported in a non-archival poster~\cite{walo_2026}.

\subsection{Method}
\begin{table*}
\begin{tabular}{l l l l}
\toprule
\textbf{ID} & \textbf{Medium (Years of Experience)} & \textbf{Experience Level} & \textbf{Platforms}\\
\midrule
P1 & Illustration (5) & Hobbyist & Instagram, TikTok  \\
P2 & Illustration (6) & Student & Instagram, Personal Portfolio Website \\
P3 & Graphic Design (25) & Professional & Instagram \\
P4 & Illustration, Game Development (6) & Student, Emerging & Instagram, Personal Portfolio Website, itch.io \\
P5 & Illustration (11) & Professional & Instagram, TikTok, Twitter/X, YouTube \\
P6 & Interactive Media (10) & Emerging & Instagram, Tumblr, Personal Portfolio Website, Bluesky \\
P7 & Printmaking and Collage (8) & Emerging & Instagram, Personal Portfolio Website \\
\bottomrule
\end{tabular}
\caption{Participants' (self-reported) demographic and background information in the formative study.}
\Description{Table summarizing the backgrounds of seven formative-study participants. Columns report participant ID, artistic medium and years of experience, experience level, and platforms used to share work. Participants span hobbyist, student, emerging, and professional experience levels and work across illustration, graphic design, game development, interactive media, printmaking, and collage. All participants used Instagram; several also used personal portfolio websites and other social platforms.}
\label{tab:participants}
\end{table*}

Our participants practiced various visual mediums and represented various career stages, as shown in Table~\ref{tab:participants}. 
Each 60-minute session combined a semi-structured interview about participants' current art distribution practices with a design-probe activity, in which we showed participants a range of existing art presentation sites (e.g., directory-based portfolios, skeuomorphic galleries, infinite-canvases) to elicit reflection on their own needs and frustrations. Sessions concluded with a moodboarding activity in FigJam\footnote{FigJam is a free online collaborative whiteboard software from Figma: https://www.figma.com/figjam/}, where the facilitator helped the participant ideate, sketch, and envision the qualities of an ideal online presentation for their artwork. We analyzed session transcripts, notes, and design artifacts using reflective thematic analysis \cite{reflecting_2019, more_2022, constructing_2006}.

\subsection{Curating an Evolving Practice}
One recurring theme in our formative study was artists' resistance to treating any single presentation as a fixed representation of their practice. Participants used different online spaces to foreground different professional, personal, social, and medium-specific needs, but these presentations also reflected changing judgments about how their work should be seen. P4, for example, regularly reordered their commission portfolio as their work improved:

\begin{quote}
``The more I take commissions, the better I think my art gets... I want to show people my best work first.'' (P4)
\end{quote}

\noindent P6 considered redesigning their website, as its current aesthetic no longer reflected how they understood their practice.

\begin{quote}
``I was thinking about redesigning my website pretty soon to embody that a little more... I just need to lean into the messiness.'' (P6)
\end{quote}

Online presentations often fail to keep pace with artists' evolving practice. Sites fell out of step with participants' current work, priorities, or aesthetic sensibilities, and the effort of restructuring an existing presentation made such changes hard to sustain. These accounts suggest that self-curation is not a one-time act of arranging finished work, but an ongoing process of reconsidering what to foreground, group, and contextualize as one's practice changes.

\subsection{Dimensions of Presentation Design}
Our formative sessions also surfaced recurring tensions in how participants evaluate presentation formats. One tension involved \textit{visibility and friction}: participants wanted more deliberate, engaging encounters than typically experienced when presenting their work in a standardized feed, but recognized that adding more interactive friction could come at the cost of accessibility and reach. P2 captured this tension directly, noting that artists ``want everything to be visible [in a] unique way,'' but that ``nothing is easier to look at than \ldots a line of images.''

A second tension concerned how strongly a digital presentation can evoke a material object, medium, or place. Several participants were drawn to situated forms that gave their work a physical sense of context---P1 imagined sketches laid across a desk, P7 pieces hanging from a clothesline, and P6 a desktop-like environment---describing such presentations as more ``intimate'' and ``intentional'' than standardized layouts.

(These tensions---visibility--friction and abstract--skeuomorphic---informed the seeded set of presentation directions we built into Mylar (\S\ref{sec:content-presentation}).)

\subsection{Design Goals for Sustaining Self-Curation}

Our formative study suggested that self-curation is less about arriving at one ideal presentation, and more about continually envisioning, and re-envisioning, how a growing and evolving body of work might be framed, encountered, and understood. This resonates with a core commitment of malleable design: that interfaces should remain open to end-user reconfiguration rather than converging on one fixed form~\cite{malleable_overview_2025}. We translated this commitment into three design goals for Mylar:

\begin{itemize}
\item \textit{DG1: Portfolios as Epistemic Artifacts.} Rather than treating generated portfolios as finished deliverables, Mylar should let artists treat them as provisional objects to think with: concrete enough to respond to, but low-commitment enough to revise or discard.
\item \textit{DG2: Multiple Presentations for Comparison and Reflection.} Artists should be able to encounter the same body of work through multiple arrangements, structures, and modes of presentation, keeping alternatives available alongside one another to compare what each foregrounds, suppresses, or suggests.
\item \textit{DG3: Continuity Across Curatorial Exploration.} Exploring a new direction should not require abandoning earlier ones. Artists should be able to reshape existing presentations and carry useful structures, interactions, or qualities from one direction into another.
\end{itemize}

Together, these goals reframe portfolio generation from producing a single presentation toward sustaining an exploratory practice of self-curation, in which arrangements can be instantiated, compared, reshaped, and repurposed as artists work through how they want their practice to be seen.

\section{Mylar System Design}
\label{sec:system}
\begin{figure}
    \centering
    \includegraphics[width=\linewidth]{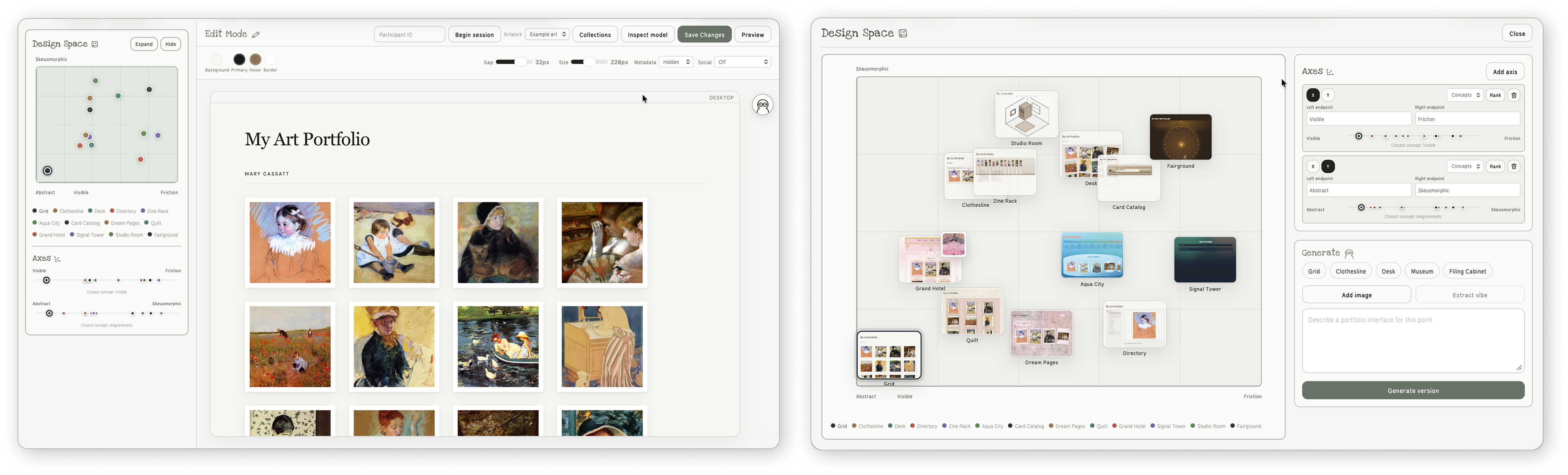}
    \caption{Left: The Mylar workspace pairs a portfolio with editing controls and a compact design space. Right: The expanded design space supports comparing directions, defining custom axes, and generating new portfolio possibilities.}
    \Description{Two screenshots show Mylar's workspace and expanded design space. Left, the workspace displays a portfolio preview alongside a compact two-dimensional design space, where points represent alternative portfolio directions positioned along abstract--skeuomorphic and visible--friction axes. Right, the expanded design space replaces the portfolio preview with a larger map of thumbnail previews, including directions such as Grid, Clothesline, Studio Room, Fairground, and Aqua City. Controls at the right allow artists to define or modify axes and generate a new portfolio.}
    \label{fig:system}
\end{figure}

Mylar is a web-based environment for exploring how artwork might be presented online. Rather than guiding an artist toward one portfolio, it keeps multiple portfolio interfaces available as provisional directions. Artists can materialize directions through language, visual references, or a design space, then reshape them through direct manipulation or natural-language edits. 

Figure~\ref{fig:system} shows Mylar’s two primary views: the workspace, which pairs a live portfolio preview with editing controls and a compact design-space panel, and the expanded design space, where artists can compare directions, define axes, and generate new possibilities. Together, these views support a recurring curatorial loop between focusing on a direction and situating it among alternatives.

Mylar's workflow embodies our three design goals. Decoupling content from presentation (\S\ref{sec:content-presentation}) keeps new directions inexpensive and provisional to produce (DG1). The design-space overview (\S\ref{sec:design-space}) lets artists hold presentations side by side for comparison and reflection (DG2). And by retaining alternatives as inputs to later prompts and preserving them through direct and AI-assisted editing (\S\ref{sec:design-space}, \S\ref{sec:generating}, \S\ref{sec:reshaping}), earlier explorations remain available rather than disappearing once superseded, sustaining continuity across curatorial exploration (DG3).

\subsection{Decoupling Artwork from Presentation} 
\label{sec:content-presentation}

\begin{figure}
    \centering
    \includegraphics[width=\linewidth]{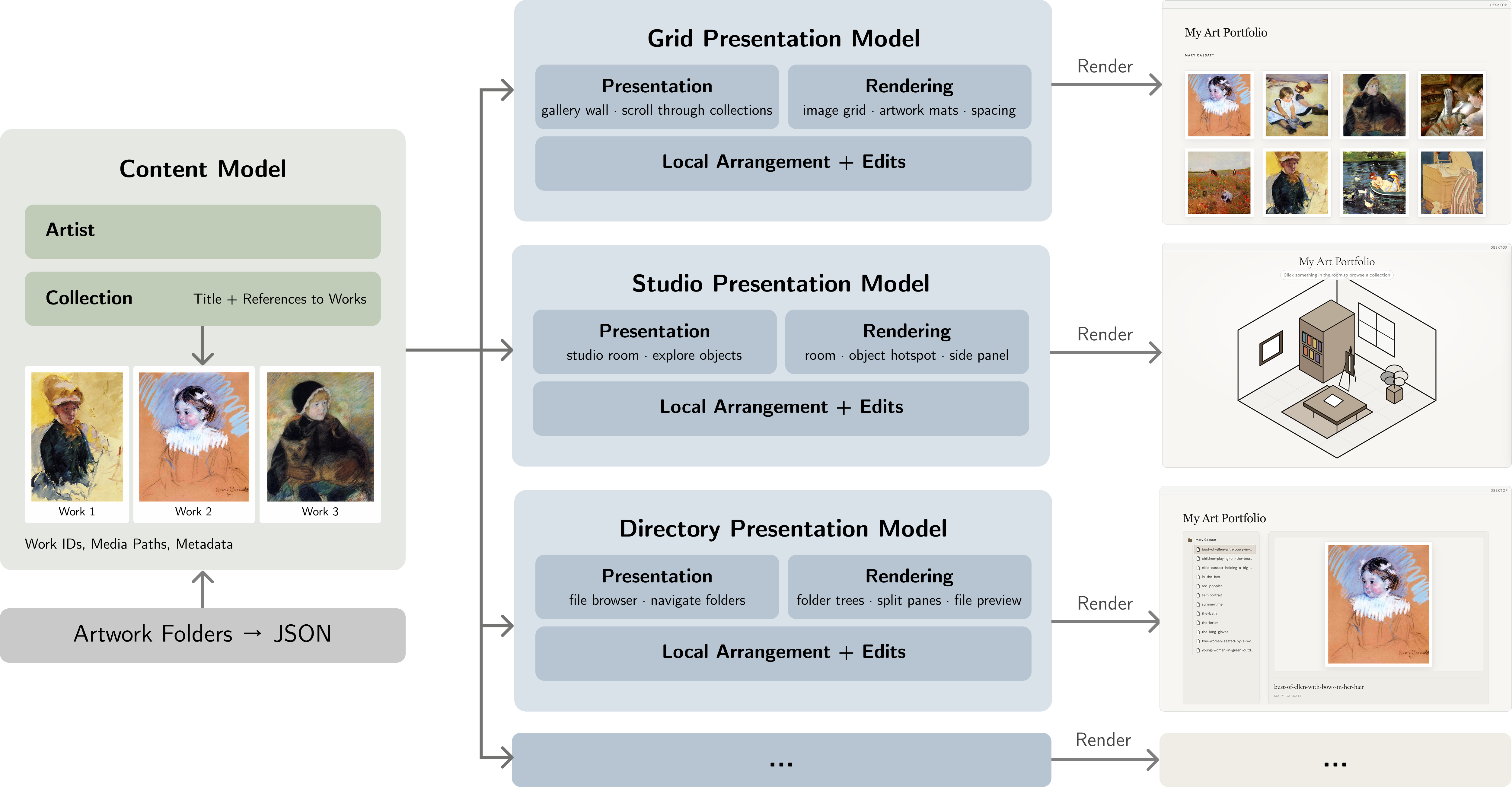}
    \caption{A single shared content model feeds multiple presentation models, each with its own organization, rendering behavior, and local edits, allowing the same underlying body of work to render as distinct portfolio interfaces.}
    \Description{Diagram showing one shared content model feeding multiple presentation models and rendered portfolio interfaces. On the left, artwork folders are parsed into a content model containing artist information, collections, and works with identifiers, media paths, and metadata. Arrows connect this shared content to several presentation models. The Grid model specifies a gallery-style presentation and image-grid rendering; the Studio model specifies a room-based presentation with object hotspots; and the Directory model specifies a file-browser presentation with folders and a file preview. Each model also contains its own local arrangement and edits. On the right, the three models render the same underlying artwork as visually distinct grid, studio-room, and directory portfolio interfaces. Ellipses indicate that additional presentation models and interfaces can be added.}
    \label{fig:decoupling}
\end{figure}

Mylar represents a portfolio as two related but independent layers---a \textit{content model} and one or more \textit{presentation models} (Figure~\ref{fig:decoupling})---so that the same body of work can be held open to more than one account of itself. 

The content model describes the artist, collections, works, and associated metadata. In the study, artists supplied a folder of their work where subfolders became collections and media files became works. Because collection labels, ordering, and membership are attributes of a given presentation rather than of the artwork itself, artists could reorganize how their work was grouped or sequenced without touching the underlying files.

A presentation model describes the layout, metaphor, navigation, disclosure, components, and rendering behavior. Participants began with thirteen seeded portfolio directions: Grid, Clothesline, Desk, Directory, Zine Rack, Aqua City, Card Catalog, Dream Pages, Quilt, Grand Hotel, Signal Tower, Studio Room, and Fairground. The set was deliberately distributed across visibility--friction and abstract--skeuomorphic dimensions identified in our formative study, ranging from conventional image-grid presentation~\cite{grid_2017} to interfaces organized through material and real-world metaphors~\cite{metaphors_1994, skeuomorph_2014}, and from immediately legible presentations to those requiring more active exploration~\cite{friction_2016}. At the implementation level, Grid, Clothesline, Desk, Directory, and Studio Room used hand-authored reusable renderers; the other eight were pre-generated presentation bundles, identical in format to portfolios artists could generate themselves.

Because presentation is fully decoupled from content, no direction is a commitment: artists can switch, create, or discard presentations freely, without duplicating or endangering their underlying work.

\subsection{Making Alternatives Commensurable}
\label{sec:design-space}
Comparing presentations meaningfully requires more than a list of past directions, but a way to see how alternatives relate to one another. Mylar retains each portfolio as a point in a two-dimensional design space, making otherwise disparate directions commensurable along shared customizable axes rather than encountering them one at a time. Its expanded view renders live thumbnail previews of all directions simultaneously, allowing artists to scan alternatives together before opening one in the main preview. The default \textit{visibility--friction} and \textit{abstract--skeuomorphic} axes foreground how much effort encountering the art requires and how strongly the interface evokes a material object or world.

Artists can create dimensions meaningful to their own practice using two verbal concepts or two images as endpoints. Mylar uses a language model to place portfolios along the axis and describe intermediate regions. Artists can correct placements by dragging directions, revising the system's interpretation. Any two axes can be mapped onto the overview, while others remain available as one-dimensional views.

The map is both a record and a generation input, letting existing alternatives feed forward into new ones rather than being left behind. Clicking a location in the design space produces an editable natural-language design brief based on its position and nearby portfolios. Artists can thus request ``more like this, but in this direction'' without naming every property, carrying qualities of previous alternatives forward rather than starting curatorial exploration over with each new direction.

\subsection{Materializing Directions from Partial Intentions}
\label{sec:generating}
\begin{figure}
    \centering
    \includegraphics[width=\linewidth]{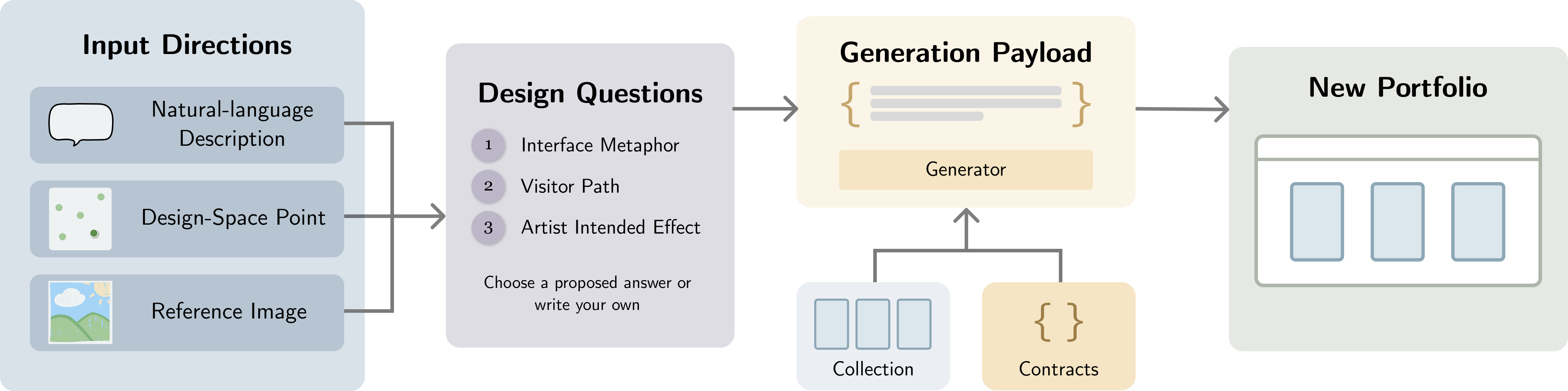}
    \caption{Mylar's portfolio generation pipeline. Artists begin from a natural-language description, a selected design-space point, or a reference image. Mylar may then ask design questions before combining the resulting direction with the artist's collection and Mylar's content and presentation contracts to generate a new portfolio.}
    \Description{Flow diagram of Mylar's portfolio generation pipeline. Three possible input directions---a natural-language description, a selected design-space point, or a reference image---feed into optional design questions about interface metaphor, visitor path, and the artist's intended effect. The resulting direction forms a generation payload, which is combined with the artist's collection and Mylar's content and presentation contracts and passed to the generator. The output is a new portfolio interface.}
    \label{fig:generation}
\end{figure}

Self-curation often begins before an artist knows exactly how they want their work to be seen, so generation cannot presume a fully-formed vision. Artists can generate from a natural-language description, a selected design-space point, or a reference image (Figure~\ref{fig:generation}). For image-guided generation, Mylar extracts palette, texture, materials, composition, typography, motifs, and interaction mood. The image supplies design direction rather than content, allowing references that are not themselves websites.

Mylar may then ask up to three design questions about the interface metaphor, visitor path, and artist's intended effect. Artists select a proposed answer or write their own, surfacing consequential ambiguities without requiring a complete specification.

The generator receives the request, answers, selected design-space position, collection summary, and Mylar's content and presentation contracts. It returns a presentation model with styling, rendering behavior, theme tokens, and optional decorative assets. Generated interfaces render works through model-bound components and expose them for later editing. New portfolios are added rather than replacing existing ones: earlier alternatives and source artwork remain unchanged, while the growing set remains available for comparison and reuse.

\subsection{Reshaping a Direction}
\label{sec:reshaping}
\begin{figure*}
    \centering
    \includegraphics[width=\textwidth]{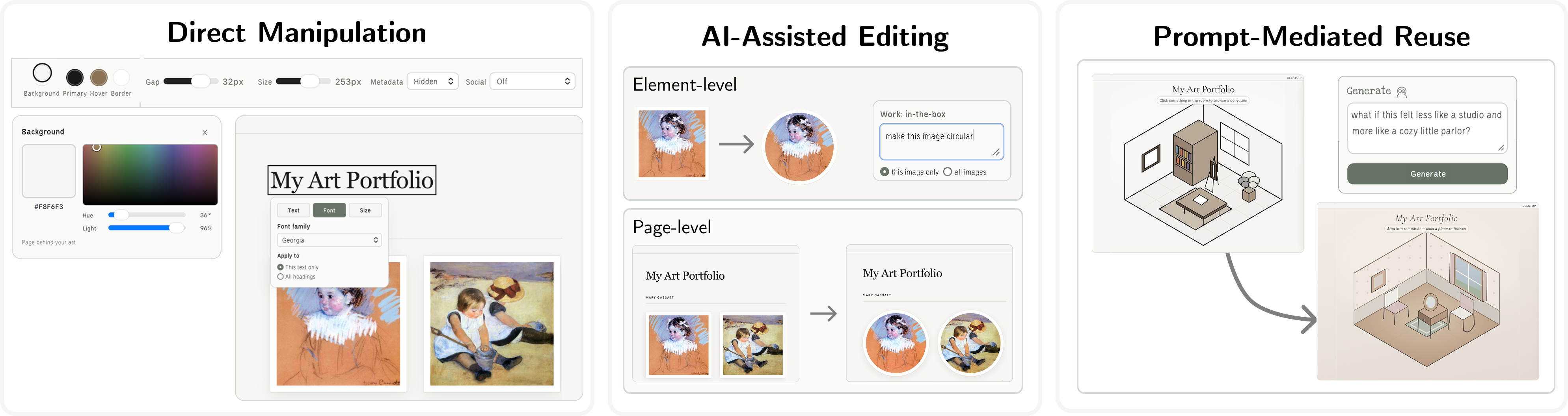}
    \caption{Reshaping a direction in Mylar. Artists can directly manipulate presentation properties, use natural-language assistance for element- or page-level edits, or reimagine an existing direction through generation. Together, these mechanisms let a presentation remain provisional while preserving useful material as artists move toward new interpretations.}
    \Description{Three panels illustrate ways artists can reshape a portfolio direction in Mylar. Left, direct-manipulation controls adjust presentation properties such as color palette, spacing, artwork size, and typography within the portfolio preview. Center, AI-assisted editing operates at two scopes: an element-level request changes a selected artwork from rectangular to circular, while a page-level edit changes the treatment of multiple artworks across the portfolio. Right, prompt-mediated reuse begins from a studio-room portfolio and a natural-language request to make it feel less like a studio and more like a cozy parlor, producing a newly generated room-based presentation that carries the earlier direction forward in altered form.}
    \label{fig:reshaping}
\end{figure*}

A generated direction makes one interpretation of the work concrete, creating something the artist can then revise rather than merely imagine. Mylar supports this reshaping through direct manipulation, AI-assisted editing, and prompt-mediated reuse (Figure~\ref{fig:reshaping}).

Controls adjust palette, spacing, artwork size, metadata placement, and direction-specific collection organization. Artists can also edit portfolio and collection titles directly.

For changes easier to express than manipulate, artists select an element and invoke a situated assistant, grounding requests such as changing one artwork's shape, restyling headings, or hiding a collection. A page-level assistant accepts broader requests about visual language, decoration, navigation, or layout. Together, these two scopes let artists reshape a direction through natural language rather than code, whether the change is local or sweeping. Both situated and page-level edits can be previewed and undone before being applied. Edits affect only the current direction, leaving source artwork and alternatives intact.

Mylar supports \textit{prompt-mediated reuse}: artists identify a structure, interaction, effect, or quality in one direction and describe it in a prompt for another. The model then re-instantiates that idea in the target direction. 

\subsection{System Architecture and Implementation}

Mylar is a browser-based editor with a model-driven JavaScript rendering layer. Its pipeline is: artwork folders are deterministically parsed into a shared JSON content model; an artist's inputs and a summary of that model form the generation context; the model returns an interface bundle; and Mylar validates and renders that bundle in a sandboxed preview. Each generated bundle carries five fields: a \texttt{presentation} object recording the interface metaphor, components, navigation, and layout behavior; scoped \texttt{css}; a \texttt{renderScript} that instantiates the interface from the shared content model; a set of \texttt{theme} tokens defining its visual language; and an optional array of decorative SVG \texttt{assets}.

Generated renderers must iterate over every collection and work, create artwork and collection elements through shared helper functions, and respect global size and spacing variables. These constraints keep generated outputs connected to Mylar's editing tools rather than producing standalone webpages.

Mylar attaches semantic identifiers for portfolio, collection, and work entities to rendered DOM elements. Direct edits address these bindings, while natural-language edits are converted into typed operations such as \textit{stylePatch}, \textit{elementStylePatch}, \textit{collectionVisibility}, or \textit{spacing}. Operations are schema-validated and restricted to allowlisted properties before application; the model does not receive unrestricted access to the page. Page-level structural changes instead return a revised bundle, which passes the same validation pipeline.

The study deployment used Claude Sonnet 4.6 for design-axis ranking, clarification questions, interface generation, and edit interpretation, and GPT-4.1 mini for extracting design tokens from reference images. Deterministic code handles content parsing, rendering helpers, state updates, and validation while model-generated output is limited to rankings, questions, structured edit operations, and interface bundles.

Generated renderers execute in opaque-origin sandboxed iframes. Network requests, persistent storage, cookies, parent-window access, external resources, and dynamic code are disabled. Participant sessions persist the generated directions, selected direction, axes, content arrangements, and direction-specific theme and style overrides. 

\section{Discovery-Oriented Evaluation Study}
\begin{table*}
\centering
\footnotesize
\begin{threeparttable}
\begin{tabularx}{\textwidth}{@{}l X l X l@{}}
\toprule
\textbf{ID} & \textbf{Medium (Years of Experience)} & \textbf{Experience Level} & \textbf{Platforms} & \textbf{Portfolio Creation} \\
\midrule
P1  & Illustration (4-5) & Hobbyist & TikTok & Hand-Coded\\
P2  & Illustration (10+) & Hobbyist & Instagram, Personal Portfolio Website & Non-Web/Physical\tnote{1} \\
P3  & Illustration, Visual Development (2) & Student & Instagram, Personal Portfolio Website & No-Code Platform \\
P4  & Digital Illustration (10), 3D Modeling (4), Game Design (4) & Student & Instagram, Twitter/X, Tumblr, Personal Portfolio Website & No-Code Platform \\
P5  & Interactive Digital Media (5) & Emerging Artist & Instagram, Personal Portfolio Website & Hand-Coded \\
P6  & Digital Illustration, Creative Coding (15+) & Professional & Instagram, Twitter/X, Personal Portfolio Website & Hand-Coded \\
P7  & Mixed Media (5) & Student & Instagram, Personal Portfolio Website & AI-Assisted/Vibe-Coded \\
P8  & Tangible Products, Creative Coding (6) & Student & Personal Portfolio Website, Behance & No-Code Platform \\
P9  & Digital Illustration, Mixed Media (10) & Professional & Instagram, TikTok, Twitter/X, YouTube & No-Code Platform \\
P10 & Digital Illustration, Graphic Design (5+) & Hobbyist & Instagram, Personal Portfolio Website & No-Code Platform \\
P11 & Digital Illustration (10) & Hobbyist & Instagram & Hand-Coded \\
P12 & Digital Art, 3D Animation, Traditional Illustration (5) & Hobbyist & Instagram, Personal Portfolio Website, Discord, YouTube & Hand-Coded \\
\bottomrule
\end{tabularx}
\begin{tablenotes}
\footnotesize
\item[1] P2's general technical background includes coding experience, but her personal portfolio was submitted as photographed physical artwork via Google Drive.
\end{tablenotes}
\end{threeparttable}
\caption{Participants' (self-reported) demographic and background information in the evaluative study.}
\Description{Table summarizing the backgrounds of 12 evaluation-study participants. Columns report participant ID, artistic medium and years of experience, experience level, platforms used to share work, and prior portfolio-creation approach. Participants span hobbyist, student, emerging, and professional experience levels and work across illustration, visual development, digital and mixed media, creative coding, game design, and related practices. Five participants previously created hand-coded portfolios, five used no-code platforms, one used AI-assisted coding, and one used a non-web physical portfolio.}
\label{tab:eval-participants}
\end{table*}

We recruited 12 artists to participate in a 60--90-minute discovery-oriented evaluation study, a kind of design probe~\cite{mattelmaki_design_2006} study that involves the design and instrumentation of a research system to facilitate a grounded evaluation of a broader community need or design situation.  Mylar served as an exploratory probe through which we could investigate broader questions about self-curation as a creative practice \cite{systems_2026}. Rather than asking whether Mylar could lead artists toward a single ``better'' portfolio, we followed how artists moved among multiple provisional forms to articulate intentions, test possibilities, revise what they wanted, and carry elements across directions. Through these encounters, we sought to understand how self-curation could become an ongoing, reflective practice when a portfolio could remain plural, provisional, and open to change.

\subsection{Participants}
We recruited 12 participants who identified as artists and had previously created a personal portfolio. Before the study, participants completed a background questionnaire about their primary artistic medium and years of experience, career/experience level, current online distribution practices, how they previously created their portfolio, as shown in Table~\ref{tab:eval-participants}. Participants also provided a collection of their own artwork for use during the free-exploration portion of the study. 

\subsection{Procedure}
Each session lasted 60--90 minutes and was conducted remotely over video call, with participants sharing their screen throughout. Sessions began with a semi-structured interview about participants' artistic practice, where and how they currently share their work online, and their prior experience building and maintaining a personal portfolio website. We then walked participants through a brief tutorial of Mylar, introducing the Design Space, the axes and generation panels, and the direct-manipulation and AI-assisted editing tools available in the central presentation view.

Participants then completed a 10--15 minute structured design task, in which they were asked to create a portfolio presentation for a small set of artworks and accompanying biographical material by the painter Mary Cassatt. Because the artwork was not their own, this task served primarily as a low-stakes way for participants to try out Mylar's generation and editing tools for the first time -- experimenting with the Design Space, axes, and AI-assisted generation without the weight of curating their own practice. Participants completed a short reflection form after the task.

Participants then completed a 20--30-minute free-exploration task using their own artwork, brought to the session in advance. Unlike the structured task, this portion placed no constraints on what participants chose to make, how many presentations they created, or how they used the system's generative and editing tools---our aim was to observe what artists chose to curate, revise, and carry across directions when given full agency over their own work. Participants again completed a reflection form after the task. Sessions closed with a short reflection interview asking participants what, if anything, the system helped them discover about how they wanted their work to live online, and how it felt to hold multiple presentations of the same body of work at once.

\subsection{Data Collection and Analysis}
We collected screen recordings, audio, transcripts, and responses to the background and mid-study reflection forms for all 12 sessions. Interviews were transcribed and analyzed using inductive thematic analysis \cite{thematic_analysis_2006}, with candidate themes developed from an initial set of interviews and iteratively refined across the remaining sessions.

A second pass of thematic coding focused on the free-exploration portions of each session, examining how participants engaged with generated interfaces as material through which they curated their own practice. Coding surfaced four recurring modes of engagement: \textit{articulating} ideas difficult to specify in advance, \textit{probing} possibilities by generating a direction and observing its consequences, \textit{foraging} across alternatives, and \textit{negotiating} intentions in response to a presentation's unexpected ``backtalk.'' We grounded these modes in prior accounts of reflective and epistemic engagement with materials~\cite{reflective_1983, distinguishing_1994, prototype_1997, initiative_1999, information_2007}, interpreting codes that emerged inductively from our data rather than predetermining them.

\section{Findings}

Across participants' free exploration, self-curation unfolded at two interconnected levels: in \textit{what} artists sought to make legible through presentation (\S6.1), and in \textit{how} they worked with provisional materials to arrive there (\S6.2).

\subsection{Curating what Becomes Legible}

\begin{table*}
\centering
\small
\begin{tabularx}{\textwidth}{
    @{}
    >{\raggedright\arraybackslash}p{0.31\textwidth}
    >{\raggedright\arraybackslash}p{0.18\textwidth}
    c
    X
    @{}
}
\toprule
\textbf{Curatorial Pattern} &
\textbf{Participants} &
\textbf{Coverage} &
\textbf{Mylar Mechanisms} \\
\midrule

Extended qualities of the work or practice into its presentation
&
P1, P3, P4, P5, P8, P9, P11
&
7/12
&
Text-to-portfolio generation, design-direction generation, custom axes, direct and AI-assisted editing. \\

\addlinespace

Structured relationships and attention within the body of work
&
P2, P3, P4, P6, P8, P9
&
6/12
&
Collection editing, direct editing, metadata placement, generated navigation, layout structures \\

\addlinespace

Situated the practice around different roles, purposes, or audiences
&
P2, P4, P7, P8, P9, P10
&
6/12
&
Text-to-portfolio generation, design-direction generation, parallel presentation alternatives, direct and AI-assisted editing \\

\bottomrule
\end{tabularx}

\caption{Recurring curatorial patterns observed during participants' free exploration with their own artwork. Patterns are ordered by participant coverage and could overlap within a participant. The final column identifies the Mylar mechanisms through which each pattern was enacted.}
\Description{Table of three recurring curatorial patterns observed during free exploration. Extending qualities of the work or practice into its presentation was observed for 7 of 12 participants and involved mechanisms including portfolio generation, custom axes, and direct or AI-assisted editing. Structuring relationships and attention within the body of work was observed for 6 of 12 participants and involved collection editing, metadata placement, navigation, and layout. Situating the practice around different roles, purposes, or audiences was also observed for 6 of 12 participants and involved generated alternatives and direct or AI-assisted editing.
}
\label{tab:curatorial-patterns}
\end{table*}

Across free exploration, participants made facets of their work, practice, and artistic selves legible through presentation, considering different accounts of themselves rather than collapsing them into a single portfolio. We observed three recurring ways participants did so: extending qualities of their work or practice into the presentation itself (7/12), establishing relationships and hierarchy within a heterogeneous body of work (6/12), and situating their practice around particular roles, purposes, or audiences (6/12) (Table~\ref{tab:curatorial-patterns}).

\subsubsection{Artists extended qualities of their work or practice into its presentation.}
For seven participants, self-curation involved making qualities of the work or the practice legible through the presentation itself, using visual language, structure, metaphor, or interaction to express something about how their work was made or how they understood it.

P1 used a text prompt and design direction to generate a presentation that displayed her collection of observational ink drawings along a digital city street (Figure~\ref{fig:portfolios}a), evoking the environment in which she typically draws:

\begin{quote}
``The context in which I demonstrate my work [in this portfolio] demonstrates that, like, oh yeah, like, I make my art usually really quickly while I'm walking on the street, and it's inspired by the people I see on the street, which I really like.'' (P1)
\end{quote}

She contrasted this with her existing sketchbook-oriented site, which she felt emphasized a more ``intimate storytelling nature'' of her practice.

Other participants similarly looked for correspondences between presentation and practice. P3 pursued a monochrome, manga-like direction in response to a largely black-and-white collection, while P5 constructed a custom ``normal''--``freaky'' axis before exploring glitchy, GeoCities- and game-inspired presentations that resonated with how she characterized her own artwork. Across these cases, what became legible was not only the artwork itself but qualities of the practice around it---the same collection could take on a different account of how it was made, what it felt like, or what kind of creative practice it belonged to, depending on the presentation through which it was encountered.

\subsubsection{Artists structured relationships and attention within their body of work.}
For six participants, self-curation also meant determining how different works should relate to one another---what belonged together, what should draw attention, and how viewers might move through the collection. Mylar's collections, layouts, navigation structures, and editable presentation hierarchy gave participants material through which to make these distinctions explicit.

\begin{figure}
    \centering
    \includegraphics[width=\linewidth]{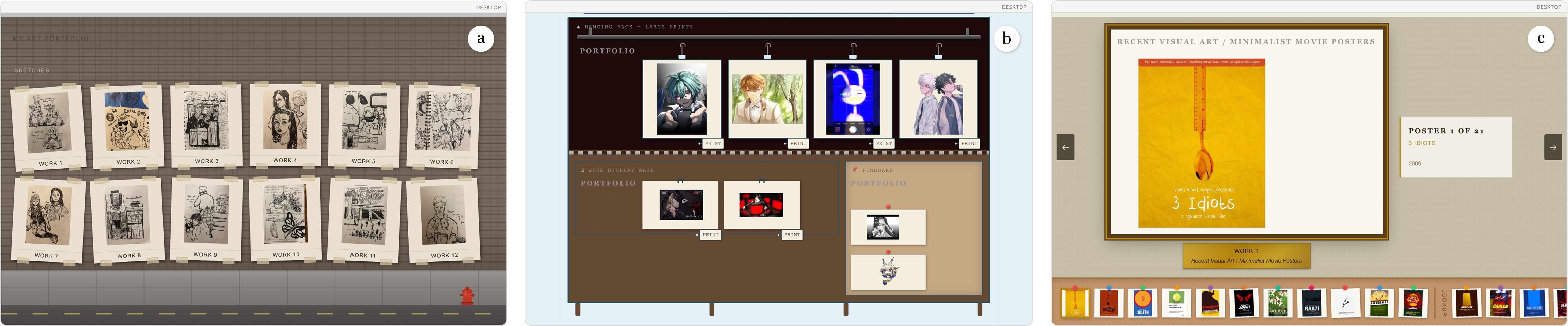}
    \caption{Three examples of self-curation through presentation. (a) P1's street-scene direction foregrounds the observational process behind her sketches. (b) P9's artist-booth-inspired direction uses print hierarchy to foreground work they identified with more personally. (c) P6's gallery-like presentation surfaces individual posters one at a time while an overview below preserves the larger collection.}
    \Description{Three portfolio interfaces created during participants' free exploration. In (a), twelve black-and-white sketches are taped to a brick wall above a sidewalk and street, forming a street-scene presentation. In (b), artworks are arranged like an artist-alley booth, with several large prints hanging across the top, smaller works grouped below, and additional pieces pinned in a side panel. In (c), a gallery-like interface presents one poster at large scale in the center with navigation arrows on either side, while a horizontal strip of thumbnail posters along the bottom provides an overview of the larger collection.}
    \label{fig:portfolios}
\end{figure}

P9 developed this most clearly while developing an artist-booth-inspired direction (Figure~\ref{fig:portfolios}b):
\begin{quote}
``This is \ldots the big, giant prints that I would \ldots have as my main thing that I want to show off. \ldots And then smaller prints \ldots [are] things you're proud of, but don't really want to invest people's attention into.'' (P9)
\end{quote}
As she continued exploring, P9 also organized commissions and other work into increasingly specific subcategories, using presentation to make different parts of their practice easier to distinguish and find.

Other participants made related moves. P2 reorganized their collection into illustrations, sketches, and oil paintings while considering which work should become a centerpiece; P6 developed a gallery-like presentation in which individual posters could be encountered one at a time while an overview at the bottom preserved the larger collection (Figure~\ref{fig:portfolios}c). P4 similarly wanted category-based navigation so visitors interested in a particular kind of work could reach it directly, while P8 imagined a query-based interface that could selectively surface projects from a broader interdisciplinary practice.

These arrangements did more than make a heterogeneous collection easier to navigate---by grouping, sequencing, enlarging, separating, or selectively surfacing work--- participants expressed judgments about relationships within their practice. Self-curation therefore involved composing an account of the body of work itself: deciding which works should speak together, which should stand apart, and what a viewer should encounter first.

\subsubsection{Artists situated their practice around different roles, purposes, or audiences.}
For six participants, what should become legible depended on the situation in which the presentation was imagined. Rather than seeking one portfolio that comprehensively represented them, participants explored accounts of their practice organized around different creative roles, purposes, and audiences.

P2 made this distinction sharply when reflecting on a generated direction:
\begin{quote}
``This just gives me the vibe of, like, I would give this to professionals for professional purposes \ldots as opposed to \ldots casual viewers or casual followers. I would see this more as a professional setting for people to review my work, to assess me.'' (P2)
\end{quote}

P9 drew a related distinction while comparing their artist-booth-inspired direction with a more conventional directory-style portfolio. The artist-booth presentation felt more closely tied to personal identification:
\begin{quote}
``This is me, like, this is, like, what I'm showing off as myself \ldots I'd probably have more, like, more personalized art on there.'' (P9)
\end{quote}
A conventional portfolio, by contrast, would contain their ``best art'' for prospective clients. For both participants, presentations organized around a professional audience and presentations organized around personal identification were not simply more and less expressive versions of the same portfolio---each foregrounded a different account of the practice, legible to a different viewer.

Participants with interdisciplinary practices encountered a related choice. P8 considered how their background in emerging technologies might become visible through a ``techie portfolio'', while P7's existing challenge of bringing design work, HCI research, and artwork into a single representation resurfaced as they considered how their work might be presented without reading as ``purely artwork''. 

Across these cases, plurality did not arise only because participants had several aesthetic preferences. Different presentations made different situated claims about who the artist was, what kind of practice they had, and what a particular visitor should encounter first. Self-curation therefore involved deciding not only \textit{how} a body of work should appear, but \textit{which account of that practice} should become salient in a given context.

Taken together, self-curation was not the task of discovering a single presentation that faithfully captured an artist's identity. Portfolio design instead became another expressive layer of the work---a means of deciding, presentation by presentation, \textit{what about my work, my practice, and myself} should come forward.

\subsection{Modes of Material Engagement}
Having traced what became legible through self-curation, we now turn to the moment-to-moment work of getting there. Self-curation unfolded through artists' ongoing engagement with the provisional material Mylar made available at every stage---existing portfolio examples, design-space axes, prompts, design questions, generated portfolios, and subsequent edits. Treating these representations as epistemic artifacts to think with, artists moved through four recurring modes of engagement: articulating, probing, negotiating, and foraging, shifting fluidly between these modes across a design trajectory as their sense of what they wanted continued to take shape.

\subsubsection{Participants articulated curatorial intentions by giving partial ideas provisional form.}
Articulation occurred when participants had some sense of what they wanted to express, but not yet a concrete account of what that intention would mean in presentation. A mood, metaphor, or desired quality could be available in language while remaining materially underspecified; instantiating it made consequences of that intention visible enough to inspect and respond to.

P9, for example, entered free exploration with the idea of an artist-alley booth, but without a clear sense of how that metaphor should organize their portfolio. After seeing it realized, they reflected:

\begin{quote}
``I had the main idea \ldots but I didn't quite have \ldots the execution. Seeing this in front of me made it easier to adjust.'' (P9)
\end{quote}

The generated version gave the concept enough form for questions of hierarchy, placement, and visual treatment to become concrete curatorial decisions about how they wanted to be seen. P8 began from an even less specified quality, wondering whether they wanted ``kind of like a techie portfolio'' while immediately adding, ``I don't know what that looks like.'' For both, articulation meant turning a quality they could name about themselves into a presentation they could encounter, judge, and elaborate.

By contrast, P12 entered generation with a comparatively specific vision and, once it materialized, immediately recognized it as ``exactly what I wanted.'' Across these cases, articulation varied in how much remained unresolved, but shared the same movement: an intention about how one wanted to be presented, given enough material form to be seen, judged, and further shaped. Provisional form made a self-presentation actionable without requiring the artist to already know exactly who that presentation should show them to be.

\subsubsection{Participants probed to test how their work might be read, before committing to a reading.}
Rather than trying to realize a particular outcome, artists sometimes engaged Mylar's representations to learn what a presentation might reveal about them before deciding what they wanted to say. What distinguished probing was not iteration on output quality, but a willingness to leave a direction underspecified so that the representation could surface how the artist's work---and by extension, themselves---might be read, before they committed to a particular reading. 

P5 made this stake explicit. Rather than generating a presentation, she rearranged Mylar's design-space dimensions around terms of her own choosing, replacing its defaults with ``beautiful'' and ``ugly'' because she was ``curious what your AI ranks as ugly.'' Her probing was not directed at what the system could produce, but at how it would position her portfolios along an interpretive axis she had set herself---a way of inspecting the space of possible readings before deciding which one she wanted to occupy.

P4 and P10 probed toward the same end from the opposite direction, using the system's generative range rather than its interpretation of existing work. P4 described this most directly:

\begin{quote}
``I want to see what it'll come up with, and what it's capable of, and then once I have a visual of what it can do \ldots I can adjust the prompt, or add details that I didn't think to add. \ldots It's more so just testing the waters before I start tweaking it more specifically.'' (P4)
\end{quote}

P10 removed detail from a prompt for a similar reason, saying he was ``sort of curious what it'll make,'' in the hope that a less specified request might produce ``something that I could work with more'' as a starting point. In both cases, under-specification was not a failure to communicate an existing intention; it was a strategy for eliciting material against which a self-presentation could take shape.

P12 offered an instructive contrast. They entered exploration with a comparatively specific website already in mind and quickly recognized a generated direction as ``exactly what I wanted.'' Reflecting on the broader exploratory interface, they distinguished between situations in which an artist still needed to ``figure out a starting point'' and their own case, where having already formed one made much of that surrounding exploration feel like ``clutter.'' Their account suggests that probing's value was contingent on how settled an artist's sense of self-presentation already was: when artists already knew how they wanted to be read, direct realization and editing could be preferable; when they did not, an unsettled representation---one that could still be interpreted many ways---was what made probing worth doing at all.

\subsubsection{Participants negotiated their presentations, revising a self-presentation already taking shape.}

Negotiation occurred when a provisional presentation pushed against a curatorial direction an artist had begun to form, prompting them to reconsider not only its concrete details but the direction itself. Unlike articulation, where a representation gave form to an intention, negotiation unfolded as a back-and-forth: artists encountered what a representation made of an emerging idea, responded to it, and sometimes carried a revised intention forward.

P3's free exploration made this movement especially visible. Looking across his own collection, he noticed that many works were black-and-white but had differently colored backgrounds. He initially reasoned toward a monochrome, manga-inspired presentation as a way of bringing them together:

\begin{quote}
``Maybe a comic book style, let me think. \ldots Let's just say manga. Comic. Let's just try that. \ldots And then I'll organize it from there, but we'll just start with that for now, and just see how that goes.'' (P3)
\end{quote}

As Mylar's design questions introduced possibilities around that direction, however, P3 began reconsidering what he had initially set out to make, noting that the process was ``trying to expand the thought process'' even as it complicated his original plan. He later reflected that the result ``did not express the original idea, but that's okay, because, it changed''---describing the value of the encounter more simply: ``it showed me things, and then I could be like, oh, I do want that.''

P3's direction therefore did not change in a single moment. An initial interpretation of his collection was given provisional form; that form introduced alternatives and consequences to respond to; and what he recognized in those encounters became part of what he wanted next. Negotiation names this recursive movement between curatorial intention and presentation. In self-curation, the artist was not simply correcting an artifact toward a stable idea of how their work should appear---the artifact could participate in changing that idea.

\subsubsection{Artists fished and foraged across portfolios for design material.}
Foraging involved treating portfolios as sources of selectively reusable material rather than as coherent candidates to keep. Artists moved among directions looking for interactions, layouts, structures, or effects that could be carried elsewhere.

\begin{quote}
``I was able to take the best of different portfolios and \ldots get inspiration [and] meld them into one.'' (P8)
\end{quote}

P5 described the generated presentations as useful ``from components, yes, but as a whole, probably not.'' P10 enacted this materially across two directions by preferring the structure of a simpler grid portfolio, but preserving an interaction effect lifted from a generated portfolio. P2 likewise evaluated a generated direction selectively, noting, ``I do like some aspects of it, and I'll probably take some inspiration from this,'' even while identifying other aspects she did not want to retain.

Across these cases, self-presentation could be assembled across portfolios, as artists gathered and recombined fragments from multiple directions rather than settling on a single candidate to keep. What persisted across exploration, then, was not necessarily a whole portfolio, but the particular qualities artists recognized as worth carrying forward---the self being assembled from parts, not simply selected from wholes.

\section{Discussion}
We discuss what these findings suggest for designing systems that support self-curation as an evolving, reflexive practice (\S7.1, \S7.2), and for using malleable prototyping as a method within formative design research itself (\S7.3).

\subsection{Designing for Plural Accounts of a Living Practice}

\S6.1 showed that participants' presentations were not always competing drafts of one eventual portfolio, but could instead make different accounts of the same practice legible at once. This finding invites a different way of designing for self-curation. Many online art platforms organize customization around a single canonical profile, grid, feed, or portfolio: artists may restyle or reorganize that representation, but alternatives ultimately give way to the version that remains. Systems for self-curation might instead support artists in holding several situated presentations in parallel, without requiring each new account to supersede the last.

P2 imagined one concrete form this plurality might take. Reflecting on how easily she could move among presentations in Mylar, she wondered whether her own website might similarly hold ``multiple sites'' that depicted her art differently for different purposes or settings. Rather than requiring a visitor to encounter only the presentation she had chosen as canonical, she imagined letting them move among these views themselves, ``expand[ing] the viewer's own idea of me'' so that she was not ``encapsulated to only one website.'' A personal portfolio might, for example, expose several presentations of a shared body of work---through different views, tabs, or entry points---each foregrounding a different role, context, or way of encountering the practice.

Supporting such experiences has architectural consequences. Treating an artist's body of work and its presentation as separable but recombinable layers---as Mylar does by binding multiple presentation models to a shared content model---allows these views to coexist without requiring artists to duplicate the work itself. Plurality therefore need not mean maintaining several independent websites: a shared collection could instead render through multiple situated presentations, with artists deciding which views remain private, which are published, and how visitors move among them. This suggests an opportunity to make plurality itself visible and navigable within creative systems, rather than treating multiplicity only as a backstage authoring convenience that must eventually resolve into a single public-facing page.

\subsection{Self-Curation as Reflexive Correspondence}

Across these encounters, what mattered was not simply that artists articulated, probed, negotiated, and foraged, but that these modes acted on a living self-presentation---as participants reshaped portfolios, they were also revising what those portfolios were meant to make legible about themselves. This process departs from what Ingold critiques as a \textit{hylomorphic} model of making, in which form is conceived in advance and imposed upon material, and instead follows what Devendorf describes as \textit{morphogenetic} making---in which form emerges through ongoing correspondence between maker and material~\cite{making_2013, machine_2015, devendorf2016strange}. In contrast with perspectives that would cast Generative AI as a tool for producing high-fidelity outcomes with well-specified prompts, Mylar intentionally foregrounds a material correspondence with presentation. Rather than fully specifying some desired final outcome for the system to enact on their behalf, artists used Mylar to give partial intentions a provisional form, to evaluate what those forms made possible, and to let that encounter inform their next actions. P7 described this process as one of recognizing something in the material and building from it:

\begin{quote}
``If I just take the grid and be like, okay, I just want to change these few things\ldots I get, like, an aha, like, I like this, and I can just keep building on top of that over and over again.'' (P7)
\end{quote}

In the context of self-curation, this correspondence has a distinctively \textit{self-reflexive} quality. In conventional curation, a curator selects, arranges, and interprets work external to themselves. In self-curation, the artist is simultaneously creator, curator, and subject of the presentation. The material being arranged is their own work, and its arrangement mediates what becomes legible about their practice and artistic self. The artist is therefore not only shaping a presentation, but encountering themselves in what that presentation gives back. As \S6.1 showed, different presentations could bring different facets of the same work, practice, and artistic self into view; correspondence here could therefore reshape both the presentation and the artist's understanding of what they wanted it to make legible.

\subsection{Malleable Prototyping as a Formative Design Method}

Our study also suggests a methodological role for malleable prototyping when participants' intentions are difficult to articulate in advance. Formative methods often ask participants to describe needs, react to existing examples, or sketch speculative alternatives; historically, prototypes produced within a study session have therefore tended toward low-fidelity forms such as paper prototypes, Wizard-of-Oz interactions, or verbal and visual sketches~\cite{paper_prototyping_2003, fingers_1994, office_1984}. Mylar made it possible for participants instead to encounter interactive, personalized possibilities during the same conversation in which those possibilities were being discussed---the kind of methodological shift recent work argues is now possible once implementation is no longer the primary bottleneck to studying working systems~\cite{systems_2026}.

The value of doing so was not simply higher fidelity or faster production. Participants could encounter working, personalized presentations generated from their artwork in real time, rather than a hypothetical mockup, a researcher-operated stand-in, or existing examples pulled from elsewhere---and then genuinely revise it and move among alternatives that were both unfamiliar and recognizably theirs. P2 described this distinction directly, contrasting generation with searching for inspiration online:

\begin{quote}
``I feel like [AI] gives me, like a bit more creative leverage, as opposed to me searching for inspiration online, and I would be kind of forced, to like look at what already exists.'' (P2)
\end{quote}

Participants could therefore respond to concrete, personalized possibilities that exceeded what they had specified in advance, using those encounters to refine, redirect, or newly articulate what they wanted their presentation to make possible. Malleable prototyping may therefore be especially useful when a formative study concerns intentions that are themselves emerging: rather than requiring participants to report stable requirements, researchers can observe how those requirements take shape through encounters with provisional artifacts.

This opportunity also introduces a methodological risk. If generated artifacts can participate in forming participants' intentions, they cannot be treated as neutral probes of preferences that already existed. The possibilities a participant encounters depend on the examples, axes, prompts, model behavior, and aesthetic tendencies built into the system; these choices may make some directions easier to imagine than others. In our study, for example, participants' exploration was necessarily shaped by Mylar's seeded presentations and by the particular interpretations produced by its generative models. The same responsiveness that allows a prototype to ``talk back'' can therefore also steer the space within which a participant comes to understand what they want.

We see malleable prototyping as promising not because it provides more direct access to participants' latent preferences, but because it makes the formation of those preferences observable and discussable. Used reflexively, such systems can extend formative design from eliciting requirements toward jointly exploring how requirements take shape---while requiring researchers to remain attentive to their own role, and the model's role, in constructing the possibilities through which participants are invited to think.

\section{Limitations and Future Work}
Our findings are grounded in a single 60–90-minute session per participant, and reflect what became visible to us within that window rather than how self-curation unfolds across the longer timescales of an evolving practice. Several participants explicitly connected Mylar's provisional presentations to changes they anticipated making to their real portfolios later; we were not able to observe whether artists carried anything forward once the session ended, whether generated directions continued to influence how they thought about their work, or whether early enthusiasm for a direction persisted after the immediacy of the study wore off. Future work might follow artists across a longitudinal deployment to examine how provisional presentations accumulate, disappear, or return over time, and whether artists revisit earlier directions as their practices, audiences, and curatorial priorities change.

Our participants primarily worked in visual, digital media, and Mylar itself centers the presentation of artwork through web-based portfolio interfaces. Practices organized around performance, sound, installation, temporality, or other forms that resist discrete digital representation may involve different forms of self-curation. Future work could extend malleable presentation beyond image-centered portfolios to explore what self-curation means when the work itself is durational, spatial, interactive, or otherwise difficult to contain within a conventional webpage. Our findings should therefore be understood as an account grounded in the kinds of artistic practices and presentation contexts represented in this study, rather than as an exhaustive characterization of self-curation across artistic practice.

\section{Conclusion}
\begin{quote}
``...our sense of artistry in any field is precisely the ability to move, in more and less articulable ways, effectively through the circumstances in which one finds oneself.''\begin{flushright}---Lucy Suchman~\cite{politics_1997}\end{flushright}

\end{quote}

We set out to explore self-curation as a creative practice: how artists explore and express their artistic identity through the presentation of their own work. We observed how 12 artists used Mylar to generate, compare, and reshape provisional presentations of their artwork collections, treating these presentations as material for design reflection. Across free exploration, we observed artists deciding which qualities, relationships, and audiences they sought to make legible through presentation---and doing so not through a single specifying act, but through ongoing articulation, probing, negotiation, and foraging across provisional forms.

This account reframes self-curation as reflexive rather than merely expressive: artists did not use presentation to communicate an already-settled sense of their practice, but encountered themselves anew in what their presentations gave back. A generated direction could talk back, revealing a quality an artist wanted before they knew they wanted it, or unsettling a plan they had already committed to. Self-curation, in this sense, is not a problem to be solved once and settled, but an evolving practice of deciding, presentation, by presentation, what about one's work, one's practice, and oneself should come forward.

We hope this account, along with Mylar itself, offers both a vocabulary for understanding self-curation as a creative practice and a design direction for systems that support artists in holding multiple, evolving accounts of who  they may be through the presentation of their work.







\bibliographystyle{ACM-Reference-Format}
\bibliography{references}



\end{document}